\documentclass[a4paper,12pt]{article}
\usepackage{amssymb,latexsym}

\makeatletter \@addtoreset{equation}{section} \makeatother

\begin{document}

\begin{titlepage}

\thispagestyle{empty}

\begin{flushright}
\hfill{HU-EP-03/28} \\
\hfill{hep-th/0306088}
\end{flushright}

\vspace{35pt}

\begin{center}{ \LARGE{\bf
BPS Action and Superpotential \\[4mm]
for Heterotic String \\[4mm]
Compactifications with Fluxes}}

\vspace{60pt}

{\bf G. L. Cardoso, G. Curio, G. Dall'Agata and D. L\"ust}

\vspace{15pt}

{\it  Humboldt Universit\"at zu Berlin,
Institut f\"ur Physik,\\
Newtonstrasse 15, D-12489 Berlin Adlershof , Germany}\\[1mm] {E-mail:
gcardoso,curio,dallagat,luest@physik.hu-berlin.de}

\vspace{50pt}

{ABSTRACT}

\end{center}

We consider ${\cal N} =1$ compactifications to four dimensions of  heterotic string
theory in the presence of fluxes.
We show that up to order ${\cal O}\left(\alpha^\prime{}^{2}\right)$
the associated action
can be written as a sum of squares of BPS-like quantities.
In this way we prove that the equations of motion are solved
by backgrounds which fulfill
the supersymmetry conditions and the Bianchi
identities.
We also argue for the expression of the related superpotential and
discuss the radial modulus stabilization for a class of examples.

\end{titlepage}

\newpage

\baselineskip 6 mm

\section{Introduction}

One of the main problems of string theory  is the large vacuum
degeneracy which appears in the effective theory following from
compactification.
This spoils part of its predictive power and poses the problem of
finding some mechanism to remove this degeneracy.
A possible way out is provided by considering flux compactifications
\cite{Strominger:1986uh}--
\cite{Berg:2003ri}.
If expectation values are assigned to Neveu--Schwarz and/or
Ramond--Ramond form--fields, a potential depending on the moduli is
generated in the four--dimensional effective theory and, in addition,
there is a backreaction on the geometry, which generically turns the
internal space into a non--K\"ahler manifold.
The appearance of a potential is important because it may lead to a
fixing of some or all of the moduli, giving us some understanding
about vacuum selection.
It may also lead to a spontaneous breaking of supersymmetry.

The possibility of introducing background fluxes was already
considered in the early studies of supersymmetric compactifications of
heterotic string theory
\cite{Candelas:1985en,Strominger:1986uh,deWit:1987xg}, but
compactifications with background fluxes never received as much
attention as those on Calabi--Yau spaces.
The main problem with implementing such a scenario is the existence of
certain no--go theorems which forbid compactifications with fluxes
using arguments following from the analysis of the equations of motion
\cite{deWit:1987xg,Maldacena:2000mw} (see also \cite{Ivanov:2000fg}
for a different approach).
These powerful arguments rule out even non--supersymmetric regular
solutions and allow only for the wellknown purely geometric
compactifications.
These no--go theorems, however, do not apply when singular sources are
added to the theory or higher derivative terms are present in the
action.
This has led to the study of many compactification examples in type
IIA/B in the presence of D--branes and orientifold planes
\cite{Dasgupta:1999ss,Kachru:2002he,Frey:2002hf,
Kachru:2002sk,Tripathy:2002qw,Blumenhagen:2003vr,Cascales:2003zp,Berg:2003ri}.
 In the heterotic context, on the other hand, the anomaly cancellation
requirement leads to a modification of the Bianchi identity of the
Neveu--Schwarz three--form and hence to the presence of higher
derivative curvature terms in the action.
Therefore, and in contrast to type II string theory, the heterotic
theory has built into it an interesting mechanism for avoiding the
above no--go theorems.
Since the heterotic action and the associated supersymmetry rules do
not depend on the introduction of sources, the analysis that can be
performed in search for flux compactifications has general validity
and the resulting classification does not depend on specific choices
in the initial setup.

The first analysis of the heterotic theory with fluxes
\cite{Candelas:1985en,Strominger:1986uh,deWit:1987xg} led to a set of
necessary and sufficient conditions for obtaining
supersymmetric solutions to first order in the $\alpha'$ expansion.
Despite the formal simplicity of the resulting geometrical conditions,
explicit solutions preserving the minimal amount of supersymmetry are
still lacking today, after more than fifteen years.
The only examples discussed in the literature so far were obtained by
applying duality transformations to type II or $M$--theory solutions
\cite{Dasgupta:1999ss,Becker:2002sx,Kachru:2002sk,Tripathy:2002qw,Becker:2003yv}.
However, it has not been shown explicitly that these configurations
solve the heterotic Bianchi identity for the Neveu--Schwarz
three--form field, and although these solutions may be exact solutions
to the full heterotic theory, their intrinsic non--perturbativeness
(at least in $\alpha^\prime$) makes it unlikely that they will satisfy
the geometrical conditions obtained in \cite{Strominger:1986uh}, since
the latter, which were derived by truncating the theory to first order
in $\alpha^{\prime}$, receive higher order corrections in $\alpha'$.

A different approach to explicitly construct and classify new examples
of heterotic backgrounds in the presence of fluxes is given by the use
of the torsion classification for $SU(3)$ structures on
six--dimensional manifolds \cite{chiossi}.
The use of the structure group of the tangent bundle provides an
organizing principle for solutions with fluxes similar to the notion
of holonomy group for the purely geometric compactifications.  Such a
tool can lead to an easy classification and construction of solutions
to the geometric conditions coming from the supersymmetry 
rules\footnote{Group structures as a tool to classify and construct 
new solutions to supergravity and string theory were first used in 
\cite{Gauntlett:2002sc} and then also in 
\cite{Kaste:2002xs,Gurrieri:2002wz,Cardoso:2002hd,Gauntlett:2002fz,Goldstein:2002pg,
Kaste:2003dh,Gauntlett:2003cy,Kaste:2003zd} 
in connection with flux compactifications.}
\cite{Gauntlett:2002sc,Cardoso:2002hd}.
Unfortunately, in order to really generate a complete solution of the
heterotic theory one also needs to solve the Bianchi identities for
the various fields, and this has to be worked out explicitly on a case
by case basis, in contrast to what happens for Calabi--Yau
backgrounds.

Solving the residual supersymmetry equations of
\cite{Strominger:1986uh} does not immediately guarantee that also the
equations of motion are satisfied.
In this note we address this issue by showing that the action of
the heterotic theory (truncated to second order in the $\alpha^\prime$
expansion) can be rewritten as a sum of BPS--like squares.
This then implies that solutions to the equations of motion can be
found by setting these BPS--like squares to zero.
An interesting outcome of this rewriting is that the conditions
appearing in the squares are precisely the geometric conditions
discussed in \cite{Strominger:1986uh,Cardoso:2002hd,Gauntlett:2003cy}.
Therefore, the heterotic equations of motion (which have contributions
from higher curvature terms) do not impose additional constraints on
the solutions constructed by solving the supersymmetry equations of
\cite{Strominger:1986uh} and the Bianchi identities for the various
fields.

Integrating the action over the internal manifold yields the potential
appearing in the four--dimensional effective ${\cal N} = 1$
supergravity theory.
This potential can be rewritten in terms of a superpotential $W$ and
various $D$--terms.
The necessary requirements for having an ${\cal N} = 1$ vacuum, i.e.
$W = 0$ and $\partial W =0$, then impose certain conditions leading to
moduli stabilization.
A rigorous derivation of the superpotential for flux compactifications
requires a detailed knowledge of the moduli space of the
compactification manifolds, which is not available at present.
An educated guess for the superpotential\footnote{The superpotential 
(\ref{supo}) and its origin were also discussed recently in 
\cite{Becker:2003gq} from an alternative point of view.}, 
which we will argue for, is
given by
\begin{equation}
W = \int \left(H + \frac{i}{2} dJ\right) \wedge \Omega\,.
\label{supo}
\end{equation}
Notice that for generic flux compactifications the internal space is
not complex, i.e. $J$ is not integrable, and therefore $dJ \wedge
\Omega \neq 0$.
Since the above superpotential explicitly depends on the three--form
flux $H$, its extremisation should give rise to the torsional
constraints leading to supersymmetric configurations.
Again, a rigorous derivation requires an explicit
knowledge of the metric moduli, but we will argue that under
certain assumptions the expected torsional constraints do follow.
More precisely, the superpotential $W$ has to lead to a determination
of $H$ in terms of the deviation of the internal space from being a
Calabi--Yau manifold.
$W$ must therefore also include pieces which are purely geometrical
and which measure the non--Calabi--Yau--ness of the internal space.
This is captured by the additional piece proportional to $dJ$ in
(\ref{supo}).

In the final section of the paper we will discuss the issue of the
radial modulus stabilization for a class of examples prominently used
in the literature.
We will argue that, on generic grounds, twisted tori cannot be used as
consistent supergravity solutions of heterotic compactifications with
fluxes, at least on the basis of an order by order $\alpha^\prime$
expansion.
We will finish with some comments on higher order $\alpha^\prime$
corrections to the conditions on the geometry of the internal space.

\section{BPS rewriting of the action}

As stated in the introduction, the main results obtained so far on
heterotic compactifications with fluxes follow from the supersymmetry
equations \cite{Strominger:1986uh}.
A solution to these equations does, however, not guarantee that the
equations of motion are also satisfied.
A nice way to find solutions to the equations of motion consists in
rewriting the action in terms of a sum of squares.
In this way, by making all the squares vanish, one also ensures the
vanishing of their variation, and therefore the action is extremised.
The resulting solutions are then described in terms of simpler
equations which often coincide with those needed to obtain
supersymmetric backgrounds.
Generically, however, one cannot expect that the equations of motion
do not impose additional constraints beyond those following from the
requirement of supersymmetry.
Here we are going to show that solutions to the heterotic
supersymmetry equations do satisfy the heterotic equations of motion,
and we do this by rewriting the heterotic action in terms of squares
(for another approach leading to a similar conclusion see
\cite{Gillard:2003jh}).

In order to show this, we will first review the conditions following
from the requirement of preserving ${\cal N}=1$ supersymmetry in four
dimensions.
Then we will show how the action of the heterotic theory truncated to
second order in the $\alpha^\prime$ expansion can be rewritten in
terms of BPS--like squares, and finally we will show that from their
vanishing one recovers the geometrical conditions following from the
supersymmetry conditions.

\subsection{Supersymmetry conditions}

Necessary and sufficient conditions for ${\cal N}=1$ spacetime
supersymmetry in four dimensions from the heterotic string
supersymmetry rules to first order in $\alpha^\prime$ were derived in
\cite{Strominger:1986uh}.
The metric in the string frame was assumed to be the
warped product of a four--dimensional maximally symmetric space--time
and a six--dimensional compact manifold,
\begin{equation}
ds^2 = g_{MN} \, dx^M \otimes dx^N =
{\rm e}^{2\left(\Delta(y)-\phi(y)\right)} \left( dx^\mu \otimes dx^\nu \, \hat g_{\mu\nu}(x)
+ dy^m \otimes dy^n \, \hat g_{mn}(y)\right)\,.
\label{metric}
\end{equation}
The relevant fermionic supersymmetry rules are given by
\begin{eqnarray}
\delta \psi_{M} &=& \nabla_{M}^{-} \epsilon  \equiv
\nabla_{M}\epsilon - \frac14 H_{MNP} \,\Gamma^{NP} \epsilon \,,
\label{gravitinor}\\
\delta \chi &=& - \frac14 \,\Gamma^{MN}\epsilon
\,F_{MN}\,,
\label{gauginor}\\
\delta \lambda &=&  \nabla\!\!\!\!\slash \phi \,\epsilon+ \frac{1}{24} \,
\Gamma^{MNP} H_{MNP}\, \epsilon  \,.
\label{dilatinor}
\end{eqnarray}
The vanishing of the supersymmetry
equations forces the four--dimensional spacetime to be Minkowskian,
i.e. ${\hat g}_{\mu\nu} = \eta_{\mu\nu}$.
For what concerns the internal six--dimensional manifold,
supersymmetry requires it to be complex, of $SU(3)$ holonomy with
respect to the special connection with torsion $\nabla^{-}$ and admitting
one nowhere vanishing globally defined holomorphic $(3,0)$--form $\Omega$.
The norm of the latter must also be related to the complex structure
$J$,
\begin{equation}
\star d \star J = i  \left( \bar \partial - \partial\right)\, \log ||\Omega|| \,.
\label{ddagger}
\end{equation}
All the ten--dimensional fields are expressible in
terms of the geometry.
The dilaton $\phi$ is related to the warp factor $\Delta$
and both are determined by the norm of the holomorphic  $(3,0)$--form,
\begin{equation}
\Delta (y) = \phi (y)  = \frac{1}{8} \log ||\Omega|| + {\rm constant} \,.
\label{delom}
\end{equation}
The three--form $H$ expectation value is determined in terms of the complex
structure,\footnote{Equations (\ref{ddagger}) and (\ref{HJ}) differ by a
sign from  the ones presented in \cite{Strominger:1986uh}.}
\begin{equation}
H = \frac{i}{2} \,\left(\partial - { \bar \partial} \right) J\,.
\label{HJ}
\end{equation}
Finally,  the Yang--Mills background field strength must be a
$(1,1)$--form satisfying $F_{mn}J^{mn} = 0$.
On top of this, one must satisfy the Bianchi identities for
the various forms.

These geometric constraints can be translated into conditions on the
intrinsic torsion $\tau$ of the Levi--Civita connection of the
compactification manifold.
It has been shown \cite{Cardoso:2002hd,Gauntlett:2003cy} that the
torsion must lie in
\begin{equation}
\label{w3w4w5}
\tau \in {\cal W}_3 \oplus {\cal W}_4 \oplus {\cal W}_5\,,
\end{equation}
that the last two are related by
\begin{equation}
2\, {\cal W}_4 + {\cal W}_5 = 0 \,,
\label{w4w5}
\end{equation}
and that
\begin{equation}
{\cal W}_4\;\;\; {\rm and} \;\;\; {\cal W}_5 \;\;\; \hbox {exact and real} \;.
\label{w4w5ex}
\end{equation}
Here ${\cal W}_1 \ldots {\cal W}_5$ denote the 5 different classes of
irreducible modules in which the torsion can be decomposed.
The various classes can be read off from $dJ$ and $d\psi$, where $J$
is the two--form associated to the almost complex structure and $\psi$
is a nowhere vanishing three--form,
\begin{eqnarray}
dJ &=& \frac34 \,i\, \left( {\cal W}_1 \,\overline\psi -\overline{\cal 
W}_1\,\psi\right)  +{\cal W}_3 + J \wedge {\cal W}_4 \;, \label{dJclass}\\
d\psi &=&  {\cal W}_1 J \wedge J + J \wedge  {\cal W}_2 + \psi \wedge  
{\cal W}_5\,, \label{dpsiclass}
\end{eqnarray}
where $J \wedge {\cal W}_3 = J \wedge J \wedge {\cal W}_2 = 0$ and $\psi 
\wedge {\cal W}_3 = 0$.

\subsection{Rewriting of the action}

Let us now rewrite the action of the heterotic theory in terms of
BPS--like
squares.
The bosonic part of the Lagrangean up to second order in
$\alpha^\prime$ is given by \cite{Bergshoeff:1989de}
\begin{eqnarray}
S &=& \int d^{10}x \, \sqrt{g}\, e^{8 \phi}\left[\frac14 \, R -\frac{1}{12}
H_{MNP}H^{MNP} + 16 (\partial_M \phi)^2 \right. \nonumber\\
&&\qquad \qquad \qquad \;\;  \left. - \frac14  \alpha^\prime
\left(F_{MN}^{I} F^{I\,MN}-  R^+_{MNPQ} R^{+\,MNPQ}  \right)
\,\right]\,.
\label{action}
\end{eqnarray}
This action is written in the string frame and its fermionic
completion makes it supersymmetric using the three--form Bianchi
identity given by
\begin{equation}
dH = \alpha^\prime \left( \hbox{tr} \,  R^{+} \wedge 
R^{+}  -\hbox{tr} F \wedge F\right)\,,
\label{HBI0}
\end{equation}
where the curvature $R^{+}$ is the generalized Riemann curvature built
from the generalized connection $\nabla^{+}$ (see the appendix).
This Bianchi identity is of particular interest, because it imposes an
interplay between terms of different orders in $\alpha^\prime$.
The $\alpha^\prime$ corrections to the standard Bianchi identity for
the three--form field are needed to cancel the Lorentz and gauge
anomalies which are present in the quantum theory of type I
supergravity in ten--dimensions coupled to Yang--Mills fields, which
is the low--energy limit of the heterotic string.
These anomalies can be cancelled by the
Green-Schwarz mechanism, which implies the modification of the
three--form Bianchi identity by $\alpha^\prime$ terms proportional
to the first Pontrjagin forms for the gauge and tangent bundles.
This cancellation still works for any choice of the torsion tensor
used in the definition of the generalized Riemann curvature appearing
in the Bianchi identity \cite{Hull:1986dx}.
Of course, one does not expect all these choices to have a different
physical meaning and indeed they may be related by field redefinitions
\cite{Horowitz:1995rf}.
Once such a choice has been made, though, the supersymmetrization of
the Bianchi identity (to a given order in the $\alpha^\prime$
expansion) imposes a specific form of the associated low--energy
equations of motion and action.
The equations of motion for the bosonic fields for some of these
possible choices and for various $\alpha^\prime$ orders have been
computed using different approaches \cite{Hull:1986xn}--
\cite{Bergshoeff:1989de}.
For the purpose of rewriting the action in terms of squares of the
supersymmetry conditions discussed above, we have used the approach
presented in \cite{Bergshoeff:1989de} and rescaled the various fields
according to the conventions of \cite{Strominger:1986uh}.

In this setup the generalized connections defining the covariantly
constant supersymmetry spinor (\ref{gravitinor}) and the one used in
the $H$--Bianchi identity (\ref{HBI0}) are different.
This choice is not the same as the one previously used in
\cite{Cardoso:2002hd}, but it turns out to be convenient for the
purpose of our calculation.
As discussed above, a different choice would have been equally valid.
One would have just needed to compute again the supersymmetric
completion of the action.
The $\alpha^\prime$ corrections would have been different, possibly
containing different combinations of the three--form field than just
the one appearing in $R^{+}$.
Note that the leading higher--order terms in the Riemann curvature in
(\ref{action}) do not appear in the characteristic form of the
Gauss--Bonnet combination, though this combination may be generated by
a vielbein redefinition \cite{Bergshoeff:1989de}.

In the search for a BPS rewriting of (\ref{action}), we will now use
the metric ansatz (\ref{metric}) and require the internal space to
admit an $SU(3)$ structure.
This means that such a manifold allows for a globally defined
orthogonal almost complex structure,
\begin{equation}
{J_m}^p {J_p}^n = - \delta_m^n\, , \quad
J_{m}{}^{p} {J_n}^q g_{pq} = g_{mn}\,,
\label{ocs}
\end{equation}
and a complex three--form $\psi$, which is of $(3,0)$--type with
respect to $J$.
In order to consistently obtain that setting to zero the BPS--like
squares implies a solution to the equations of motion, we also impose
that the only degrees of freedom for the various fields are given by
expectation values on the internal space and are functions only of the
internal coordinates.

To simplify the discussion we limit ourselves to the case with
dilaton and  warp factor identified, i.e. $\phi = \Delta$, but the
generalization of the following results is straightforward.
After some manipulations, which we are going to describe in the next
subsection, the action (\ref{action}) can be written as
\begin{eqnarray}
S  = && \int d^4 x \, \sqrt{g_{4}}\, \left\{- \frac{1}{2} \int_{{\cal
M}_{6}} \, {\rm e}^{8\phi}\left(8 d\phi + \theta \right)\wedge \star \left(8 d\phi + \theta \right)
+\frac18 \int_{{\cal M}_{6}} {\rm e}^{8\phi}\, J \wedge
J \wedge {\hat R }^{ab}J_{ab} \right.\nonumber \\
& -&\frac14 \int d^{6}y \; \sqrt{g_6} \,{\rm e}^{8\phi}\,
N_{mn}{}^p \,g^{mq}g^{nr}g_{ps}\,N_{qr}{}^s
\,
\nonumber \\
&+&  \frac12 \int_{{\cal M}_{6}} {\rm e}^{8\phi}\,
\left(H + \frac12 \star {\rm e}^{-8\phi}\, d({\rm e}^{8\phi}\, J)\right) \wedge \star \left(H + \frac12 \star
{\rm e}^{-8\phi}\, d({\rm e}^{8\phi}\, J)\right) \,\nonumber\\
&-&  \frac{\alpha^\prime}{2}\int d^{6}y\; \sqrt{g_6}\,
 {\rm e}^{8\phi}\, \left[\hbox{tr} (F^{(2,0)})^{2} +
{\rm tr}(F^{(0,2)})^{2} + \frac14\,\hbox{tr} (J^{mn}F_{mn})^{2}\right] \nonumber \\
 &+& \left. \frac{\alpha^\prime}{2} \int d^{6}y\;\sqrt{g_6}
\, {\rm e}^{8\phi}\, \left[\hbox{tr} (R^{+\,(2,0)})^{2} +
\,\hbox{tr} (R^{+\,(0,2)})^{2} + \frac14 \,\hbox{tr} (J^{mn}
 R^{+}_{mn})^{2}\right]\right\}\,.
\label{finalaction}
\end{eqnarray}
In this expression the traces are taken with respect to the fiber
indices $a,b,\ldots$, whereas the Hodge type refers to the base
indices $m,n,\ldots$ of the curvatures.
The other geometrical objects appearing in the above
expression are the Lee--form
\begin{equation}
\theta \equiv J \lrcorner dJ= \frac{3}{2} J^{mn} \,
\partial_{[m}J_{np]} \, dx^p\,,
\label{eqLee}
\end{equation}
the Nijenhuis tensor
\begin{equation}
N_{mn}{}^p = {J_m}^q \partial_{[q}J_{n]}{}^p -  {J_n}^q
\partial_{[q}J_{m]}{}^p\,,
\label{eq:Nijenhuis}
\end{equation}
and the generalized curvature $\hat R$, which is constructed using the
Bismut connection built from the standard Levi--Civita connection and a
totally antisymmetric torsion $T^B$ proportional to the complex structure,
\begin{equation}
T^B_{mnp} = \frac32 \,{J_m}^{q} {J_n}^{r} 
{J_p}^{s}\partial_{[q}J_{rs]} = -\frac32 J_{[m}{}^{q} \nabla_{|q|} J_{np]}\,.
\label{eq:tor}
\end{equation}
The action (\ref{finalaction}) will now be used to find the conditions
determining the background geometry.
If, on the other hand, one also wants to consider fluctuations around
these backgrounds then (\ref{finalaction}) can be used as the expression
for the scalar potential of the effective four--dimensional theory,
\begin{equation}
S = - \int d^4 x \,\sqrt{g_4} \,V\,.
\label{eq:Vdef}
\end{equation}

The action (\ref{finalaction}) consists of a sum of squares as well as
of one linear term.
In order to have a solution of the equations of motion one sets to
zero all the squares and proves that the linear term does not
contribute to the equations of motion.
We will come back to the linear term in a moment, but first we would
like to exhibit the correspondence between the supersymmetry
conditions and the terms which are squared in the action.
The geometrical conditions resulting from the vanishing of the
BPS--like squares are the vanishing of the Nijenhuis tensor
$$N^{m}{}_{np} = 0$$ and of some components of the generalized Riemann
curvature constructed from the $\nabla^{+}$ connection,
$$R^{+\,(2,0)}=R^{+\,(0,2)}=J^{mn}R^{+}_{mn} = 0.$$
The vanishing of the Nijenhuis tensor states that the internal
manifold is complex (which means ${\cal W}_{1} = {\cal W}_{2} = 0$ in
the torsion classes language).
The conditions on the $R^{+}$ curvature can be translated into the
integrability constraints following from the vanishing of the
gravitino supersymmetry transformation (\ref{gravitinor}), which leads
to the requirement of $SU(3)$ holonomy for the $\nabla^{-}$
connection.
The proof requires the identity
\begin{equation}
R^{+}_{ab\,cd} = R^{-}_{cd\,ab} - (dH)_{abcd}\,,
\label{eq:rpm}
\end{equation}
which relates the $R^{+}$ and $R^{-}$ curvatures with the base and
fiber indices swapped.
Using this identity and the fact that $dH$ gives higher order terms
in $\alpha^\prime$ the conditions on the base indices of $R^{+}$ become
conditions on the $R^{-}$ fiber indices, to lowest order in $\alpha^\prime$,
\begin{equation}
R^{-\,(2,0)} = R^{-\,(0,2)}=J^{ab}R^{-}_{ab} = 0\,.
\label{eq:rm}
\end{equation}
These conditions precisely state that the generalized curvature $R^-$
is in the adjoint representation of $SU(3) \subset SO(6)$ and
therefore its holonomy group is contained in $SU(3)$.
We also obtain the relation between the ${\cal W}_{4}$ and ${\cal
W}_{5}$ torsion classes, expressed by the identification of the
differential of the dilaton with the Lee--form,
\begin{equation}
d\phi +\frac18 \, \theta =0\, \,,
\label{dfth}
\end{equation}
as follows from (\ref{ddagger}) and (\ref{delom}). 
The conditions in the gauge sector are also the same conditions one
gets from the supersymmetry variation of the gaugino, i.e. the gauge
field strength is of type $(1,1)$ and $J$ traceless.  
The final BPS--like square precisely yields the locking condition of
the three--form $H$ onto the almost complex structure $J$ which was
derived from the vanishing of the supersymmetry transformations.
Indeed, on a complex manifold, and using (\ref{dfth}), the following
identity holds \cite{Gauntlett:2001ur},
\begin{eqnarray}
H = -  \frac12 \star {\rm e}^{-8\phi}\,d({\rm e}^{8\phi}\,J) =
\frac12 i ( \partial- {\bar \partial }) J \;. \label{locking}
\end{eqnarray}


Let us now show that the linear term does not give any further 
condition.
The locking condition (\ref{locking}) implies the
identification of the connection $\nabla^{-}$ defining the
generalized curvature $R^{-}$ with the Bismut connection defining the
generalized curvature $\hat R$.
In this way, on the solution, the remaining linear term can also be rewritten as
\begin{equation}
\int {\rm e}^{8\phi}\, J \wedge J \wedge \hat R^{ab}J_{ab}  = 2 \int {\rm 
e}^{8\phi}\,\star J \wedge c_{1}\left(R^{-}\right)\,.
\label{eq:chernc}
\end{equation}
The Ricci two--form $c_{1}\left(R^{-}\right)$ vanishes according to 
(\ref{eq:rm}) and so does the linear term.
Now consider varying (\ref{eq:chernc}) with respect to the metric, i.e.
\begin{equation}
\delta \int {\rm e}^{8\phi}\,J \wedge J  \wedge \hat R^{ab}J_{ab} = 
2 \int {\rm e}^{8\phi}\,\star J \wedge \delta c_{1}(\hat R)\,,
\label{varchern}
\end{equation}
where we used that
$c_{1}(\hat R)=c_{1}(R^{-}) =0$ on the solution.
The Ricci two--form of the Bismut connection is the same as the Ricci form
of the Chern connection up to an exact form \cite{Alexandrov,Ivanov:2000ai},
\begin{equation}
c_1 (\hat R) = c_1 (\hbox{Chern}) -2\, d \left(\star d\star J \right) \,,
\label{eq:chern}
\end{equation}
where the Chern connection is defined as
\begin{equation}
{C^m}_{np} = {\Gamma^m}_{np} +\frac32 {J_n}^q \, \partial_{[q}J_{rp]}\,g^{rm}\,.
\label{chernconn}
\end{equation}
The variation of $c_1({\rm Chern})$, when evaluated on a manifold which is complex, yields
\begin{equation}
\left. \phantom{\int} \delta \,c_1({\rm Chern})\right|_{N_{mn}{}^{p}=0}= \frac12 
\;  d\left[ J\cdot d 
\left(g^{mn}\delta g_{mn}\right)\right]\,,
\label{eq:varc1}
\end{equation}
where $J\cdot d = dx^{m} J_m{}^n \partial_n$.  
This implies that
\begin{equation}
\left. \phantom{\int} \delta c_1 (\hat R)\right|_{N_{mn}{}^{p}=0} = d \Lambda\,,
\label{eq:delc1}
\end{equation}
for a certain one--form $\Lambda$.
Hence the variation (\ref{varchern}) is vanishing once an integration
by parts is performed,
\begin{equation}
\int {\rm e}^{8\phi}\,\star J \wedge \delta c_{1}(\hat R)= 
\int {\rm e}^{8\phi}\,\star J \wedge d\Lambda = 
- \int d \left({\rm e}^{8\phi}\,\star J\right) \wedge \Lambda = 0\,,
\label{eq:vaefi}
\end{equation}
and once one uses (\ref{dfth}).

Before proceeding with the derivation of (\ref{finalaction}) let us
briefly comment on the limiting case with constant dilaton $\phi= const.$ and 
vanishing flux $H = 0$.
The locking condition (\ref{locking}) simply becomes the
requirement for the internal manifold to be K\"ahler, imposing $dJ =
0$. 
In addition, the square involving the dilaton becomes a condition imposing 
the vanishing of the Lee form, $d\star J = 0$.
Moreover, now $\nabla^{\pm}=\nabla$ and $R^+ = R^- = R$.
Therefore the conditions on the holonomy of $\nabla^{-}$ become
conditions on the Levi--Civita connection.
The solution is obviously given by Calabi--Yau manifolds,
which are  K\"ahler and have
vanishing first Chern class.


\subsection{Details of the derivation}

We now discuss the derivation of the action (\ref{finalaction}) in
detail.  
We will see that several interesting geometrical
relations arise and that some non--trivial cancellations occur.

The first interesting and quite non--trivial relation emerges when
one tries to express the Ricci scalar of the  Levi--Civita connection
in terms of other geometric quantities of the internal manifold.
After a tedious calculation one  obtains that
\begin{equation}
\begin{array}{rcl}
\displaystyle
\int d^{6}y\;\sqrt{g_6} \, {\rm e}^{8\phi}\,R &=&\displaystyle - \int {\rm e}^{8\phi}\,\theta
\wedge \star \theta + \frac12 \,\int
 {\rm e}^{8\phi}\,
J \wedge J \wedge R^{ab}J_{ab}  - 16 \int \,
{\rm e}^{8\phi}\, \, d\phi \wedge \star
\theta\\[4mm]
&&-2  \displaystyle  \int {\rm e}^{8\phi}\, dJ^{(3,0)+(0,3)} \wedge \star
dJ^{(3,0)+(0,3)}\nonumber\\
&&- \;\,\displaystyle
\int d^{6}y\;\sqrt{g_6} \; {\rm e}^{8\phi}\,
N_{mn}{}^p \,g^{mq}g^{nr}g_{ps}\,N_{qr}{}^s \,,
\end{array}
\label{eq1}
\end{equation}
where total derivatives were dropped.
Therefore the Ricci scalar has been related to the square of
the Lee form $\theta$, the Nijenhuis tensor $N_{mn}{}^p$,
the $(3,0)+(0,3)$ part of the $dJ$ tensor,
the double $J$ trace of the Riemann tensor $R_{ab\,cd}J^{ab}J^{cd}$ 
and the mixed term between $d\phi$ and $\theta$ which we need as a 
double product in the square relating them.

An obvious manipulation which has to be performed is the rewriting of
the vector kinetic term as squares of the conditions following from
the gaugino supersymmetry transformations.
To this end we use the following identity,
\begin{eqnarray}
\int d^{6}y\;\sqrt{g}\,{\rm e}^{8\phi}\,\hbox{tr} (F_{mn}F^{mn}) =
&+& 2 \int d^{6}y\;\sqrt{g}\, {\rm e}^{8\phi}\,\left[\hbox{tr} (F^{(2,0)})^{2} +\hbox{tr} (F^{(0,2)})^{2} + \frac14
{\rm tr}(J^{mn}F_{mn})^{2}\right]\nonumber \\
&-&\displaystyle 2 \int  \,{\rm e}^{8\phi}\,J \wedge\,\hbox{tr} (F \wedge F) \,.
\label{eqF}
\end{eqnarray}
The same relation can also be applied to the higher order term in the
Riemann curvature appearing in (\ref{action}), which can be rewritten
as
\begin{equation}
\begin{array}{rcl}
&&\displaystyle\int d^{6}y\;\sqrt{g_6}  \, {\rm e}^{8\phi}\, R^{+}_{mnab}
R^{+\,mnab} = \int d^{6}y\, \sqrt{g_6} \, {\rm e}^{8\phi}\,
{\rm tr}(R^{+}_{mn}R^{+mn}) \nonumber\\[4mm]
&&\qquad \qquad \displaystyle =+ 2 \int d^{6}y\;\sqrt{g}  {\rm
e}^{8\phi}\, \left[\hbox{tr} (R^{+(2,0)})^{2} +\hbox{tr}
(R^{+(0,2)})^{2} + \frac14
{\rm tr}(J^{mn}R^+_{mn})^{2}\right] \nonumber\\
&& \; \; \;\qquad \qquad  \displaystyle -2 \int  {\rm
e}^{8\phi}\,J \wedge\hbox{tr} (R^{+} \wedge R^+) \,.
\end{array}
\label{eqR}
\end{equation}
We can now attempt to perform the first rewriting of the action
(\ref{action}) using (\ref{eq1}), (\ref{eqF})  and (\ref{eqR}),
\begin{eqnarray}
S  = && \int d^{4}x \, \sqrt{g_4} \left\{- \frac{1}{4} \int \, {\rm
e}^{8\phi}\left(8 \,d\phi + \theta \right)\wedge \star \left(8 \,d\phi +
\theta \right)  \right.\nonumber\\
&+&\frac18 \int \,{\rm e}^{8\phi}\,  J \wedge
J \wedge R^{ab}J_{ab} - \frac12\int \,{\rm e}^{8\phi}\,  \star H \wedge H
\nonumber \\
&-&\frac12 \int\,{\rm e}^{8\phi}\,  dJ^{(3,0)+(0,3)} \wedge \star
dJ^{(3,0)+(0,3)}  - \frac14 \int d^{6}y\;\sqrt{g} \,{\rm e}^{8\phi}\,
N_{mn}{}^p \,g^{mq}g^{nr}g_{ps}\,N_{qr}{}^s  \, \nonumber\\
&+&\frac{\alpha^\prime}{2} \int \,{\rm e}^{8\phi}\, J \wedge \left[-{\rm tr} (R^{+} \wedge R^{+})+
{\rm tr} (F \wedge F)\right]\nonumber\\
&-& \frac{\alpha^\prime}{2}  \int d^{6}y\;\sqrt{g_6} \,{\rm e}^{8\phi}\, \left[{\rm tr} (F^{(2,0)})^{2} +
tr(F^{(0,2)})^{2} + \frac14{\rm tr} (J^{mn}F_{mn})^{2}\right] \nonumber \\
 &+&\left. \frac{\alpha^\prime}{2}  \int d^{6}y\;\sqrt{g_6} \,{\rm e}^{8\phi}\, \left[{\rm tr} ( R^{+(2,0)})^{2} +
 {\rm tr} ( R^{+(0,2)})^{2} + \frac14{\rm tr} (J^{mn}
 R^{+}_{mn})^{2}\right]\right\}\,.
\label{action2}
\end{eqnarray}
In order to obtain the relations coming from the analysis of the supersymmetry
transformation laws it is now more natural to introduce the generalized
Riemann curvature $\hat R$ constructed from the Bismut connection.
An explicit computation shows that
\begin{eqnarray}
\int  \,{\rm e}^{8\phi}\,J \wedge J \wedge R_{ab}J^{ab} = 
&&\int \,{\rm e}^{8\phi}\,J \wedge J \wedge
{\hat R}_{ab}J^{ab} + 2 \int \,{\rm e}^{-8\phi}\,\left[d
({\rm e}^{8\phi}\,J) \wedge \star \, d ({\rm e}^{8\phi}\,J)\right]
\nonumber \\
&+& \frac12  \int d^{6}y\;
\sqrt{g_6} \,{\rm e}^{-8\phi}\, \left[\star d ({\rm e}^{8\phi}\,J) 
\right]_{mnp}\,J^{mq}\,
J^{nr}\, \left[\star d({\rm e}^{8\phi}\,J) \right]_{qr}{}^p
\nonumber \\
&-& 2  \int \, {\rm
e}^{8\phi}\left(8 \,d\phi + \theta \right)\wedge \star \left(8 \,d\phi +
\theta \right)\,,
\label{c1}
\end{eqnarray}
where we performed an integration by parts on the second term
and where we discarded total derivatives.
Using (\ref{c1}) and the three--form Bianchi identity (\ref{HBI0}), some
of the terms in (\ref{action2}) may be rearranged as
\begin{eqnarray}
&&\frac18 \int {\rm e}^{8\phi}\, J \wedge
J \wedge R^{ab}J_{ab} - \frac12 \int {\rm e}^{8\phi}\, \star H \wedge H
- \frac12 \int {\rm e}^{8\phi}\, J \wedge dH
\nonumber\\ 
&=&\frac18 \int {\rm e}^{8\phi}\,J \wedge
J \wedge {\hat R}^{ab}J_{ab}
+ \frac12 \int{\rm e}^{8\phi}\, \left(H + \frac12 \star {\rm
e}^{-8\phi}\,d({\rm e}^{8\phi}\,J)\right) \wedge \star \left(H + \frac12
\star {\rm e}^{-8\phi}\, d({\rm e}^{8\phi}\,J)\right) \nonumber\\
&-& \frac14  \int \, {\rm
e}^{8\phi}\left(8 \,d\phi + \theta \right)\wedge \star \left(8 \,d\phi +
\theta \right)
\label{c1h} \\
&+& \frac18 \int {\rm e}^{-8\phi}\,\left[d
({\rm e}^{8\phi}\,J) \wedge \star \, d ({\rm e}^{8\phi}\,J)\right] + \frac{1}{16}
\int \sqrt{g_6} \, {\rm e}^{-8\phi}\, \left[\star d ({\rm e}^{8\phi}\,J) 
\right]_{mnp}\,J^{mq}\,
J^{nr}\, \left[\star d({\rm e}^{8\phi}\,J) \right]_{qr}{}^p\,,
\nonumber
\end{eqnarray}
where we performed an integration by parts on the $J \wedge dH$ term 
in order to reproduce the square which gives the 
$H$--locking condition.
The last line of this expression is of a special form which selects a
specific projection of the three--form $dJ$.
Using the projection relations (which we review in the appendix) it
can be shown that for a given three--form $T$
\begin{equation}
4 \int
\star (T^{(3,0)}+T^{(0,3)}) \wedge (T^{(3,0)}+T^{(0,3)})
= \int \star T \wedge T - \frac12 \int d^{6}y \,\sqrt{g_6}\,
 T_{mnp} J^{mq} J^{nr} T_{qr}{}^p \,,
\label{hhT}
\end{equation}
which, upon identification of $T$ with $-\frac12{\rm e}^{-8\phi}\, \star
d({\rm e}^{8\phi}\,J)$, yields a new expression
for the last line in (\ref{c1h}).

Inserting now (\ref{c1h}) and (\ref{hhT}) into (\ref{action2}) and
applying all
due simplifications one obtains the following
expression for the action,
\begin{eqnarray}
S = &&\int d^{4}x \, \sqrt{g_4} \left\{- \frac{1}{2} \int \, {\rm
e}^{8\phi}\left(8 \,d\phi + \theta \right)\wedge \star \left(8 \,d\phi +
\theta \right)  \right.\nonumber\\
&+&\frac18 \int J \wedge
J \wedge {\hat R }^{ab}J_{ab}
-  \frac14 \int d^{6}y\;\sqrt{g} \,{\rm e}^{8\phi}\,
N_{mn}{}^p \,g^{mq}g^{nr}g_{ps}\,N_{qr}{}^s  \, \nonumber\\
&+& \frac12 \int
{\rm e}^{8\phi}\,
\left(H + \frac12 \star {\rm e}^{-8\phi}\, d({\rm e}^{8\phi}\, J)\right) \wedge \star \left(H + \frac12 \star
{\rm e}^{-8\phi}\, d({\rm e}^{8\phi}\, J)\right) \nonumber \\
&-&\frac12 \int {\rm e}^{8\phi}\, \star \left(\star d J
\right)^{(3,0)+(0,3)} \wedge \left(\star dJ\right)^{(3,0)+(0,3)}
-\frac12 \int {\rm e}^{8\phi}\,dJ^{(3,0)+(0,3)} \wedge \star
dJ^{(3,0)+(0,3)}
\nonumber\\
&-& \frac{\alpha^\prime}{2}  \int d^{6}y\;\sqrt{g_6} \,{\rm e}^{8\phi}\, \left[{\rm tr} (F^{(2,0)})^{2} +
tr(F^{(0,2)})^{2} + \frac14{\rm tr} (J^{mn}F_{mn})^{2}\right] \nonumber \\
 &+&\left. \frac{\alpha^\prime}{2}  \int d^{6}y\;\sqrt{g_6} \,{\rm e}^{8\phi}\, \left[{\rm tr} ( R^{+(2,0)})^{2} +
 {\rm tr} ( R^{+(0,2)})^{2} + \frac14{\rm tr} (J^{mn}
 R^{+}_{mn})^{2}\right]\right\}\,.
\label{action4}
\end{eqnarray}
The terms proportional to $dJ^{(3,0)+(0,3)}$
cancel out because $\star dJ^{(3,0)+(0,3)} = \left(\star
dJ\right)^{(3,0)+(0,3)}$, which follows from the projector property
\begin{equation}
P_{a}{}^i P_b{}^j P_c{}^k \epsilon_{ijkdef} =  \epsilon_{abcijk}
Q_d{}^i Q_e{}^j Q_f{}^k \,.
\label{eq:Peps}
\end{equation}
Consequently one obtains the result (\ref{finalaction}) for the
action.

\section{The superpotential and the torsional constraints}

In this section we want to address the problem of deriving
the torsional constraints and the potential of the effective ${\cal N}=
1$ four--dimensional theory in terms of a superpotential $W$.
We will not present a rigorous proof but rather
try to work in close analogy with the type IIB theory, pointing out
the various problems that arise in doing so.


In the case of type IIB compactifications with 3--form fluxes
\begin{equation}
G=H^{(3)}_R-\tau H^{(3)}_{NS}\,,
\label{GIIB}
\end{equation}
the potential of the effective theory is given by the sum of two terms
depending on the warp factor $A$ \cite{Giddings:2001yu,DeWolfe:2002nn}
\begin{equation}
\label{viib}
V={1\over 2\kappa_{10}\Im\tau}\int e^{4 A} G^+\wedge\star \overline{ G^+}
+{i\over 4\kappa_{10}\Im \tau}\int e^{4A}G\wedge \bar G\,,
\end{equation}
where the first term is an F-term whereas the second term is a topological
D-term.
In (\ref{viib}) the complex three--form $G$ was split into its self-- and
anti--selfdual parts
\begin{equation}
G=G^{+}+G^- \, ,
\end{equation}
where
\begin{equation}
\star G^+=+ i G^+\, ,\quad
\star G^-=- i G^-\, .
\end{equation}
As expected, the $F$--term of the potential can be derived from a superpotential
which is given by
\cite{Gukov:1999gr,Taylor:1999ii,Mayr:2000hh,Giddings:2001yu,DeWolfe:2002nn}
\begin{equation}
\label{supoiib}
W=\int G\wedge\Omega=\int(H^{(3)}_R-\tau H^{(3)}_{NS})\wedge\Omega\, ,
\end{equation}
and its extremization yields the conditions on the fluxes to obtain a
supersymmetric Minkowski vacuum.


Let us now try to understand how far such a derivation can be
extended to the heterotic theory with fluxes. 
We saw in the rewriting of the effective action in terms of BPS--like
squares (\ref{finalaction}) that there are several terms contributing
to the final potential.
These terms have different sources and nature and are related to both
the introduction of fluxes and to the modification of the geometry.
Of course we would like to understand the precise interpretation of
some of these terms as F-- or D--terms in the construction of the
${\cal N} = 1$ effective theory.  
As a first step in this direction we will simply consider the terms
involving the $H$--flux contribution, namely
\begin{equation}
V = -\frac12 \int
{\rm e}^{8\phi}\,
\left(H + \frac12 \star {\rm e}^{-8\phi}\, d({\rm e}^{8\phi}\, J)\right) \wedge \star \left(H + \frac12 \star
{\rm e}^{-8\phi}\, d({\rm e}^{8\phi}\, J)\right) \, .
\label{eq:I}
\end{equation}

In trying to push the analogy with the type IIB case one can
introduce the complex three--form
\begin{equation}
{\cal H} \equiv H + \frac{i}{2} \, {\rm e}^{-8\phi}\, d({\rm e}^{8\phi}\, J)\,,
\label{eq:calH}
\end{equation}
which, as we will argue below, plays the same role as the complex
three--form $G$ on the type IIB side. 
The potential $V$ given in (\ref{eq:I}) can be rewritten in terms of
the three--form $\cal H$ as
\begin{equation}
V = -\frac{1}{2}\left( ||{\rm e}^{4\phi}\,{\cal H}||^2 - i \int {\rm
e}^{8\phi}\, {\cal H} \wedge \bar{{\cal H}} \right)\,,
  \label{hetasIIBaction}
\end{equation}
where we introduced the inner product\footnote{This product is
positive definite: $\langle\alpha,\alpha\rangle \ge 0$ and
$\langle\alpha,\alpha\rangle =0 \Leftrightarrow \alpha =0$.  } of
three--forms on a complex manifold $X$
\begin{equation}
\langle\alpha_1,\alpha_2\rangle \equiv\int_X\alpha_1\wedge
\star \bar \alpha_2\,,
\label{InProdC}
\end{equation}
and the norm is simply the inner product of a form with itself.

In perfect analogy with the IIB case one can split the complex
three--form $\cal H$ in terms of its self-- and anti--selfdual parts
\begin{equation}
{\cal H} = {\cal H}^{+}+ {\cal H}^{-}\,,
\label{eq:Hsplit}
\end{equation}
where
\begin{equation}
\star {\cal H}^+=+ i {\cal H}^+\, ,\quad
\star {\cal H}^-=- i {\cal H}^-\, .
\end{equation}
After this decomposition is performed the potential
(\ref{hetasIIBaction}) further simplifies to
\begin{equation}
V = -\int {\rm e}^{8\phi}\,{\cal H}^{+} \wedge \star
\overline{\cal H^+}\,. \label{eq:I2}
\end{equation}
This is formally the same as the first term in (\ref{viib}).
Note that there is no analogue of the topological term.
This is in contrast with the IIB case, where such a term was
crucially involved in the tadpole cancellation.
This is, however,
not too surprising since the topological term was intimately
related to the presence of localized sources that do not play any
role in the case of heterotic compactifications.

At this point one would like to show that the potential (\ref{eq:I2})
can be written in the
standard ${\cal N}= 1$ form
\begin{equation}
V = {\rm e}^{\cal K}\left[g^{i\bar \jmath}D_i W \, \overline{ D_{\bar \jmath}
W} - 3|W|^{2}\right]\,,
\label{eq:n1}
\end{equation}
where $\cal K$ is the K\"ahler potential of the moduli space and
$g^{i\bar \jmath}$ is the inverse metric following from $\cal K$.
In the IIB case this followed from the fact that $e^{4A}G^{+}$ is
harmonic on the conformally rescaled internal space, which is
Calabi--Yau, and therefore one can expand it in terms of the harmonic
forms on the same manifold.
In the heterotic case, however, one is faced with two problems: the
moduli space of the internal manifold is not known and the complex
form ${\cal H}^{+}$ is neither closed nor coclosed.
For what concerns the first problem one would need to obtain a better
understanding of the geometric properties of the six--dimensional
manifolds \cite{Cardoso:2002hd} involved in obtaining supersymmetric
vacua of the heterotic theory.
Unfortunately, it is difficult to understand the general structure of
their moduli space.
An additional source of complications is given by the fact that
not all the manifolds in such classes are solutions also of
the three--form Bianchi identity (\ref{HBI0}).

Since the ${\cal H}$--form is not closed nor co--closed, one
cannot expand it into harmonic forms of the internal
space as in the IIB case \cite{Giddings:2001yu,DeWolfe:2002nn}.
The hope is that some other expansion may be performed which replaces
that of \cite{Giddings:2001yu,DeWolfe:2002nn},
and that from such an expansion one can
read off the superpotential in an analogous way.
Assuming this to be the case, a simple guess for the superpotential is given by
\begin{equation}
W = \int {\cal H} \wedge \Omega = \int \left( H + \frac{i}{2} dJ
\right) \wedge \Omega \,. \label{eq:supo}
\end{equation}
Note that the term proportional to $d \phi \wedge J$ in $\cal H$
has dropped out of (\ref{eq:supo}) due to its Hodge type.
A similar superpotential was also suggested in \cite{Becker:2003gq},
through a dualization procedure of the IIB potential.
Here we provide alternative arguments for justifying the guess
(\ref{eq:supo})  with some new motivations for the appearance of the
extra $dJ$ term.

It is interesting to notice that in (\ref{eq:supo}) the two--form potential
$B$ has been combined with the almost complex structure two--form $J$
(which is not closed in general) to give the complexified area
\begin{equation}
{\cal J}   = \,   B + \,  \frac{i}{2}\, J\,.
\end{equation}
In the K\"ahler case $d{\cal J} = dB$, so that at the
field--strength level one does not see the explicit appearance of
this complex combination. 
In the heterotic theory one also needs to extend the three--form
field--strength to include the gauge and Lorentz Chern--Simons terms,
$H=dB + \alpha^{\prime}\left(\omega_L - \omega_{YM}\right)$.  
On the other hand, supersymmetry imposes that the introduction of
$H$--fluxes must be accompanied by a
non--vanishing $dJ$ and these two are mutually non--local, i.e. $H
\wedge dJ \neq 0$.\footnote{In view of the analogy with
type IIB, we note that there as well it is not possible to have
non--trivial supersymmetric flux solutions when only one flux system,
say $H_R$, is turned on: if $e\times m$ is zero, then one can rotate
all fluxes to be electric, but $W=e_I X^I$ is not non--trivially
solvable in the $e_I$ on a non--singular Calabi--Yau space.}
Thus it seems appropriate to introduce the
combination $\cal H$ in the definition of the superpotential
(\ref{eq:supo}).


Extending further the analogy with the type IIB case, one can ask
whether the locking condition (\ref{locking}) follows from
$W = \partial W=0$.
The vanishing of the superpotential $W = 0$ implies that the $(0,3)$
part of ${\cal H}$ has to vanish.
On the other hand, in the heterotic case we do not have an explicit dilaton
dependence in the complex three--form defining the superpotential, as
was the case in the IIB theory (\ref{GIIB}).
This means that one cannot obtain the vanishing of the $(3,0)$ part of
${\cal H}$ from the variation of the superpotential with respect to
the dilaton.  
On the other hand, assuming that an overall radial modulus $\rho$ exists, the
scaling behaviour of the almost complex structure on $\rho$ is given by
\begin{equation}
dJ = (\rho +\bar \rho) dJ|_{\rho +\bar \rho=1}\,.
\label{eq:radial}
\end{equation}
Extremization of the superpotential\footnote{Note that the 
superpotential is nevertheless holomorphic in $\rho$.} with respect to $\rho$ then gives
\begin{equation}
dJ^{(0,3)}= 0\,.
\label{eq:dJ30}
\end{equation}
By reality it follows then that
\begin{equation}
dJ^{(3,0)} = dJ^{(0,3)} =0\,,\quad H^{(3,0)} = H^{(0,3)} =0\,.
\label{dJH30}
\end{equation}
It is then easy to see that this implies the vanishing of the first 
torsion class ${\cal W}_{1}= 0$ in (\ref{dJclass}).
Next, assuming that $H$ is constant under variation of the almost complex
structure moduli and that $\partial_{i} \Omega  = k_{i}\Omega +
\chi_{i}$, where $\chi_{i}$ are a set of $(2,1)$ forms and $k_{i}$
are constants over the internal manifold, one further obtains that the
$(1,2)$ part of $\cal H$ is vanishing.
Altogether, one finds that most of the components of the
complex three--form ${\cal H}$ have to vanish,
\begin{equation}
{\cal H}^{(3,0)} = {\cal H}^{(0,3)} ={\cal H}^{(1,2)}=0\,.
\label{eq:van}
\end{equation}
On a complex manifold, the last of these conditions yields the
locking of the three--form flux onto the geometry, namely
\begin{equation}
{\cal H}^{(1,2)} = H^{(1,2)} + \frac{i}{2}\,\bar
\partial\, J = 0\,,
\label{eq:HdbarJ}
\end{equation}
and since $H$ is a real form, the desired relation follows,
\begin{equation}
H = \frac{i}{2} \left( \partial - \bar \partial\right) J\,.
\label{eq:HdJ}
\end{equation}


An analogous situation arises also in type IIA compactifications
with RR 2--form flux. 
In this case the backreaction of the fluxes induces a strong
deformation of the geometry such that ${\cal M}_6$ will not be any
more complex or K\"ahler.
This can be seen performing \cite{Kachru:2002sk} a mirror
transformation starting from IIB models with NS 3-form flux, or
lifting type IIA with 2-form flux to M-theory on a $G_2$ holonomy manifold (see
the discussion in the appendix of \cite{Cardoso:2002hd} and
\cite{Kaste:2002xs,Kaste:2003dh,Kaste:2003zd}).
The result is that ${\cal M}_6$ is characterized by
non--vanishing torsion in ${\cal W}_2^+\,$.
So the type IIA space ${\cal M}_6$ is an almost--K\"ahler manifold.
This space has a non-closed 3-form $\Omega$, i.e. $d\Omega\neq 0$.
As first discussed in \cite{Vafa:2000wi} (see also the brief
discussion in \cite{Curio:2001dz}) and further elaborated on in
\cite{Gurrieri:2002wz}, $d\Omega$ provides a purely geometrical term
to the type IIA superpotential, a so--called NS 4-form flux, and the
type IIA superpotential then takes the form
\begin{equation}
W=\int(H^{(2)}_R\wedge J+d\Omega)\wedge J\, .
\end{equation}
Like for the term $dJ\wedge\Omega$ in the heterotic superpotential
(\ref{supo}), the deviation of ${\cal M}_6$ from being a
Calabi--Yau space is measured in type IIA by $d\Omega\wedge J$
(note that by partial integration these two terms agree), and
space--time supersymmetry links together $H^{(2)}_R$ with $d
\Omega$.

\section{Comments on the radial modulus and higher order $\alpha^\prime$ corrections}

In the quest for explicit supersymmetric solutions of the heterotic
theory with fluxes the five torsion classes
for $SU(3)$ structures defined in \cite{chiossi} are an important tool.
As shown in \cite{Cardoso:2002hd},
given the information about the intrinsic torsion of the allowed
manifolds, it is very easy to provide entire classes of
candidate solutions by simply consulting the mathematical literature.
In the same way, new classes of solutions can be constructed upon analysis
of the conditions
(\ref{w3w4w5})--(\ref{w4w5ex}) \cite{Goldstein:2002pg}.
However, as pointed out in \cite{Cardoso:2002hd}, it does not seem
possible to also solve the $H$ Bianchi identity for a given general
class as was the case for Calabi--Yau manifolds.
This means that, in order to decide whether a certain manifold
provides a consistent supersymmetric heterotic background, one needs
to explicitly solve the modified three--form Bianchi identity
(\ref{HBI0}) on a case by case basis.

An interesting issue which arises for these solutions is the
stabilization of the overall size modulus.
In all the known examples of IIB compactifications the resulting
effective four--dimensional models were of no--scale type
\cite{Taylor:1999ii,DallAgata:2001zh,
Kachru:2002he,Louis:2002ny,Ferrara:2002bt,DAuria:2002th,Kachru:2002sk}
and the overall size modulus remained free even after the introduction
of fluxes.
This cannot happen in the heterotic compactifications because of the
modified Bianchi identity for the three--form \cite{Candelas:1985en,Gauntlett:2003cy}.
As we will now show, the various terms appearing in equation (\ref{HBI0})
generically scale differently with respect to this parameter and
therefore it needs to be fixed in
order to obtain a consistent solution.
Given a generic metric with overall size modulus $t\equiv \sqrt{\rho 
+\bar \rho}$, i.e. $ds^2 =
t^2 (...)$, the two--form associated to the complex structure $J$
scales with $J \to t^2 J$.
Since for supersymmetric backgrounds the three--form $H$ is related to
$J$ by (\ref{locking}), its exterior derivative
scales accordingly, i.e.
\begin{equation}
dH \to t^2 dH\,.
\label{eq:sHscal}
\end{equation}
On the other hand it is straightforward to check the following scaling
relations for  the curvature,
\begin{equation}
R^a{}_{bcd} \sim t^0 \quad \,,  \quad {\rm tr} R \wedge R \sim t^0\,.
\end{equation}
Consequently, the generic solution of
the $H$ Bianchi identity (\ref{HBI0})
imposes a second order equation in the
parameter $t$, which will generically result in $t$ becoming
fixed.

This feature raises the question of the validity of the
supergravity approximation we have used.
The quadratic equation fixing the value of the overall size modulus
depends on $\alpha^\prime$  and generically leads to
compactification manifolds with a
characteristic radius of $\sqrt{\alpha^\prime}$.
This, of course, implies a breakdown of the supergravity approximation
and therefore one should not trust the generic solution found in this
way.
A possible way out \cite{Becker:2003gq}
is to allow a dependence of $t$ on the flux density, which enters in
the determination of the background geometry.
For instance,
the authors of \cite{Becker:2003gq} argue that
the overall size modulus of their solution is given by
$
t^{6} = \alpha^{\prime} |f|^{2}\,,
$
where $f$ is the flux density.
Since in principle $f$ can be taken to be arbitrarily large,
the overall size of the manifold can still be large.
As appealing as this feature may be, it does unfortunately not
guarantee that the supergravity approximation can be trusted, as we
will now show for a special class of manifolds.

An interesting subclass of solutions of the geometrical constraints
(\ref{w3w4w5})--(\ref{w4w5ex}) is given by manifolds which are
products of tori and twisted tori (included in the class of the
nilmanifolds in \cite{Cardoso:2002hd}) or by torus fibration of
Calabi--Yau two--folds, where the fiber is twisted
\cite{Becker:2002sx,Cardoso:2002hd,Goldstein:2002pg,Becker:2003yv,Gauntlett:2003cy,Becker:2003gq}.
A simple instance of such spaces is given by a metric of the form
\begin{equation}
ds^2 = t^2 \, [ d s_{4}^{2}(z,\bar z) + (dv + N \sum_j f^j(z,\bar z) dz^j)
\, (d\bar v + N \sum_j \bar f^j(z,\bar z) d\bar z^j)] \;,
\label{twistedmetric}
\end{equation}
where we use dimensionless coordinates.
In this equation the twist in the metric is associated to a
deformation parameter $N$, which is related to the flux.
Using (\ref{eq:HdJ}) one finds that
\begin{equation}
H \sim N \, t^2 \,\sum_k df^{k} \wedge dz^k \wedge (d\bar v + N  \sum_j\bar f^jd\bar z^j)\,,
\label{flux}
\end{equation}
from which it follows that
\begin{equation}
dH \sim N^2 \, t^2 \,\sum_{j,k} df^k\wedge d\bar f^j \wedge dz^k \wedge d\bar
z^j\,.
\label{dff}
\end{equation}
An analogous computation for the ${\rm tr} R \wedge
R$ terms yields
\begin{equation}
{\rm tr} R \wedge R \sim t^0 \, N^4\,  \sum_{j,k} df^k\wedge d\bar f^j \wedge dz^k \wedge d\bar
z^j\;.
\end{equation}
Inserting this into the $H$--Bianchi identity
(\ref{HBI0}) shows that
the overall size is fixed not only in terms of $\alpha^\prime$, but
also in terms of the flux as
\begin{equation}
t \sim N \sqrt{ \alpha^\prime}\;.
\label{tn}
\end{equation}
Therefore a large value of $N$ results in a large value of $t$.  
A large value of $t$, however, does not guarantee that the
supergravity approximation can be trusted.
Indeed, when evaluating the Ricci scalar on the solution, one finds
that it is of order $1/\alpha^\prime$, i.e.
\begin{equation}
R \sim t^{-2} \, N^2 \sim \frac{1}{\alpha^\prime}\,,
\label{r}
\end{equation}
where we used (\ref{tn}). 
Thus not only the size but also the curvature scales with the flux.
Since each term in the effective action, when evaluated on the
solution, is of order $1/\alpha^{\prime}$, one should retain the
entire tower of $\alpha^{\prime}$ terms in the action.
The reason why this happens for twisted tori metrics is because the
$\alpha^\prime \to 0$ limit is singular for these backgrounds.

One may thus be tempted to instead consider manifolds which admit a
smooth $\alpha^\prime \to 0$ limit.
However, in this case the resulting geometry is simply a deformation
of the underlying Calabi--Yau manifold, and it has been shown that
there is no moduli fixing \cite{Gillard:2003jh} at all.
Another possibility for achieving a consistent moduli fixing in
heterotic compactifications with fluxes is of course to consider
solutions which still do not admit a smooth $\alpha^\prime \to 0$
limit, but which do not fall into the class described above.


To conclude this discussion, let us try to understand how the
manifolds described by the constraints (\ref{w3w4w5})--(\ref{w4w5ex})
will be modified by higher order terms in $\alpha^\prime$.

The only way to obtain the exact background geometry
for flux compactifications is by a detailed knowledge of the
corresponding  string $\sigma$--model.
As a first approximation, one considers the
related supergravity theory for which the  $\alpha^\prime$  corrections
can be computed  by supersymmetry requirement.
These terms follow from the modification of the $H$--Bianchi identity
due to the Green--Schwarz anomaly cancellation mechanism.
This modification breaks the supersymmetry of the standard type I
supergravity coupled to super Yang--Mills in ten dimensions.
To restore it one has to add new terms both to the supersymmetry
rules and to the action.
This can be done either order by order in $\alpha^\prime$ 
\cite{Bellucci:1988ff,Bergshoeff:1989de},
or by directly including all order corrections giving rise to
non--linear curvature
parameterizations \cite{Bonora:1988xn,DAuria:1988yv,Raciti:1989je}.
An important drawback of both these approaches is the existence of
ambiguities in the resulting theories which can only be fixed by the
string computations.
For instance, it is known that at the quartic level the effective
action obtained from loop $\sigma$--model computations
contains $R^{4}$ invariants whose coefficient goes with
$\zeta(3)$.
This term does not appear explicitly in any of the supergravity computations as
it is not required by supersymmetry.
The hope is that the corrections obtained in this way may be a
consistent truncation of the full theory.
In any case, they are indicative of the final structure of the
supersymmetry transformation rules and are helpful in providing some
understanding of what will happen to the
torsional constraints which were derived
to first order in $\alpha^\prime$.

The supersymmetry transformations of the gravitino and dilatino do
not depend anylonger on the same combination of the heterotic
three--form $H$, as was the case to $O(\alpha^\prime)$,
but on two different tensors,
\begin{eqnarray}
\delta \psi_{M} &=& \nabla_{M}\epsilon - \frac{1}{36}\Gamma_{M}
\Gamma^{NPQ} T_{NPQ} \,\epsilon\,, \\
\delta \lambda &=& -2 \nabla\!\!\!\!\slash \phi \,\epsilon +
\Gamma^{MNP} Z_{MNP}\, \epsilon \,.
\end{eqnarray}
The torsion tensor $T$ also appear in the definition of the $H$
Bianchi identity,
\begin{equation}
dH = \alpha^\prime\left( \hbox{tr} \, \widetilde R \wedge
\widetilde R  -\hbox{tr} F \wedge F\right)\,,
\end{equation}
since the definition of the generalized curvature $\widetilde R$ includes the torsion $T$.
The difference with the order zero analysis is that now $T$ is not
simply identified with (a conformally rescaled) $H$, but it receives
non--linear higher order corrections,
\begin{equation}
e^{\frac43 \phi}T_{MNP} = -3 H_{MNP}  + \alpha^\prime W_{MNP}\,.
\label{eq:TH}
\end{equation}
The same happens to the $Z$ tensor which is defined as
\begin{equation}
e^{\frac43 \phi} Z_{MNP} = \frac16 T_{MNP}+  \alpha^\prime X_{MNP}\,.
\label{eq:ZT}
\end{equation}
Here both $W$ and $X$ are three--forms depending non--linearly on
$T$ and $\widetilde R$ (The relations are discussed in a certain Weyl frame, 
but the outcome is frame independent).
Obviously it is almost impossible to discuss the resulting geometrical
constraints, but it is also clear that the relations
(\ref{w3w4w5})--(\ref{w4w5ex}) will be modified.
In particular it is not guaranteed that the internal geometry will
stay complex, since this condition was crucially dependent on the fact
that both the gravitino and dilatino supersymmetry variation depended
on the same three--form $H$, whereas here we have two different
tensors.
Moreover, even if the geometry still has $SU(3)$ holonomy with respect
to the generalized connection with torsion $T$, we do not expect that
(\ref{locking}) will persist to higher orders in $\alpha^{\prime}$.
The derivation of this locking condition depended once more on the
peculiar structure of the supersymmetry transformations to lowest
order in $\alpha^{\prime}$, with the same type of structure entering
in the gravitino and dilatino rules.
The only relation which will remain valid in the presence of higher
order $\alpha^{\prime}$ corrections is the identification of the
Nijenhuis tensor with the $(3,0)+(0,3)$ parts of the torsion tensor,
\begin{equation}
 N_{mnp}= 4 \, T^{(3,0)+(0,3)}_{mnp} \,.
\label{eq:TN}
\end{equation}

\bigskip

\noindent
{\bf Acknowledgments}

\medskip

We would like to thank M.~de Roo,  K.~Lechner, I.~Pesando and N.~Prezas
for valuable discussions as well as K.~Dasgupta, B.~K\"ors, A.~Krause,
D.~Martelli, G.~Papadopoulos and D.~Waldram for useful
correspondence.
The work of G.L.C.\ is supported by the DFG.
Work supported in part by the European Community's Human Potential
Programme under contract HPRN--CT--2000--00131 Quantum Spacetime.

\appendix

\setcounter{equation}{0}

\section{Appendix}

The definition for the Ricci scalar $\widetilde R^a{}_{bac} \, g^{bc}$
follows from the Riemann curvature,
\begin{equation}
{\widetilde R}^a{}_{bcd} = 2\partial_{[c} \widetilde\Gamma^a{}_{d]b} +2 
\widetilde
\Gamma^a{}_{[c|m} \widetilde \Gamma^m{}_{d]b}
\label{eqriemann}
\end{equation}
computed from the connection with torsion $T$, \begin{equation}
\widetilde \Gamma^a{}_{bc} = {\Gamma^a}_{bc}+T^{a}{}_{bc}\,.
\label{eqgamma}
\end{equation}
Moreover the covariant derivative is defined as
\begin{equation}
\widetilde\nabla_{a}{J_{b}}^c = \partial_{a}{J_{b}}^c - \widetilde
\Gamma^s{}_{ab} {J_{s}}^c + \widetilde \Gamma^c{}_{as} {J_{b}}^s\,.
\label{eqcovd}
\end{equation}
For a totally antisymmetric torsion tensor it also follows that
\begin{equation}
\widetilde{R}_{ij} = R_{ij} - T_{ilm}T_j{}^{lm}- \nabla^{k}T_{kij}\,,
\label{eqricci}
\end{equation}
and
\begin{equation}
\widetilde{R} = R - T_{abc}T^{abc}\,.
\label{scal}
\end{equation}

The torsion $T$ for the heterotic theory will be $\pm H$. The
associated spin connection with torsion is then given by ${\widetilde
\omega}_m\,^{ab} = \omega_m\,^{ab} \mp H_m\,^{ab}$,
 and this is how one
distinguishes the corresponding connections $\nabla^{\mp}$ and
curvatures $R^{\mp}$.

In the paper we considered almost--complex Riemannian manifolds admitting an
$SU(3)$ structure.
Using the almost complex structure, one can define the projector tensors
\begin{equation}
{P_a}^b = \frac12 \left(\delta_a^b - {\rm i} {J_a}^b \right)\,,
\quad{Q_a}^b = \frac12 \left(\delta_a^b + {\rm i} {J_a}^b \right)\,,
\label{eq:projectors}
\end{equation}
which define the $(1,0)$ and $(0,1)$ projected components of a
generic form.
For instance, we can split the components of a three--form, say $T$, as
follows,
\begin{eqnarray}
T^{(3,0)}_{abc} & = & {P_a}^d {P_b}^e {P_c}^f T_{def}\,, \nonumber\\
T^{(2,1)}_{abc}  & = &  3 {P_a}^d {P_b}^e {Q_c}^f T_{def} \,,
\label{eq:Tcomp}\\
T^{(1,2)}_{abc} & = & 3 {P_a}^d {Q_b}^e {Q_c}^f T_{def} \,, \nonumber\\
T^{(0,3)}_{abc} & = & {Q_a}^d {Q_b}^e {Q_c}^f T_{def} \,, \nonumber
\end{eqnarray}
and the total decomposition of the three--form $T$ according to type
is unique
\begin{equation}
T_{abc}=T^{(3,0)}_{abc}+T^{(2,1)}_{abc} +T^{(1,2)}_{abc} +T^{(0,3)}_{abc}
\,.
\label{eq:Tdec}
\end{equation}
The same decomposition can be done for any other generic $p$--form.
Exterior differentiation will preserve the type only for complex
manifolds.

Some very useful relations which appear when one considers only ``real''
combinations are
\begin{eqnarray}
T^{(3,0)}_{abc}+T^{(0,3)}_{abc} & = & \frac14 \left(T_{abc} - 3
J_{[a}{}^p J_b{}^q T_{c]pq} \right) \,,\label{eq:H30}\\
T^{(2,1)}_{abc}+T^{(1,2)}_{abc} & = & \frac34 \left(T_{abc} +
J_{[a}{}^p J_b{}^q T_{c]pq} \right)  \,.\label{eq:H21}
\end{eqnarray}
If one in particular considers the square of (\ref{eq:H30}) a further
simplification arises
\begin{equation}
4\left(T^{(3,0)}_{abc}+T^{(0,3)}_{abc}\right)^{2} =  T_{abc}T^{abc} - 3 T_{abc} J^{aq}
J^{bs} T_{qs}{}^c\,.
\label{eq:H2}
\end{equation}
and hence
\begin{eqnarray}
\!\!\!\!\!\!\!\!\!\!\!\! 4 \int
\star (T^{(3,0)}+T^{(0,3)}) \wedge (T^{(3,0)}+T^{(0,3)})
&=& \frac{1}{6} \int d^6 y\,\sqrt{g} \, \left( T_{abc}T^{abc} - 3 T_{abc} J^{aq}
J^{bs} T_{qs}{}^c \right) \nonumber\\
&=& \int \star T \wedge T - \frac12 \int d^6 y\,\sqrt{g}\,
 T_{abc} J^{aq} J^{bs} T_{qs}{}^c \,,
\label{hh}
\end{eqnarray}
which is very useful in identifying some of the expressions appearing
in the rewriting of the action in terms of BPS--like squares.

Imaginary (anti--)selfduality is defined according to
\begin{eqnarray}
T^{\pm} & = & \frac12 \left(T\mp i \star T\right) \,,\label{eq:TSD}\\
\star T^{\pm} & = & \pm i \, T^{\pm}\,. \label{eq:Tpm}
\end{eqnarray}

The anomaly cancellation implies a modification of the Bianchi
identity for the three--form $H$,
\begin{equation}
dH = \alpha^\prime \left( p_1(F) -p_1(R)\right)\,,
\label{eq:p1}
\end{equation}
where $p_1(F)$ and $p_{1}(R)$ are the first Pontrjagin forms for the
gauge and tangent bundles respectively.
In terms of the Riemann and Yang--Mills curvatures this equation reads
\begin{equation}
dH = \alpha^\prime\left( \hbox{tr} \, \widetilde R \wedge
\widetilde R  -\hbox{tr} F \wedge F\right)\,,
\label{dH}
\end{equation}
where the trace on the gauge indices (and similarly for the
generalized Riemann curvature $\widetilde R$) is taken as
\begin{equation}
{\rm tr}\left(F\wedge F\right)= F^{ab}\wedge F_{ab}\,,
\label{traces}
\end{equation}
and
\begin{equation}
F^{ab} = F^{I} (T^{I})^{ab}\,, \hbox{  with  } \quad {\rm
tr}\left(T^{I}T^{J}\right) = \delta^{IJ}\,.
\label{eq:FT}
\end{equation}



\providecommand{\href}[2]{#2}\begingroup\raggedright\endgroup

\end{document}